\documentclass[aps,twocolumn,showpacs,preprintnumbers,prb,superscriptaddress]{revtex4-1}
\usepackage{graphicx}
\usepackage{amssymb}
\usepackage{amsmath}
\usepackage[toc,page]{appendix}
\usepackage{color}
\newcommand{\R}{\mathbf{r}}

\newcommand{\Q}{\mathbf{q}}

\newcommand{\be}{\begin{equation}}
\newcommand{\ee}{\end{equation}}
\newcommand{\bea}{\begin{eqnarray}}
\newcommand{\eea}{\end{eqnarray}}
\newcommand{\bean}{\begin{eqnarray*}}
\newcommand{\eean}{\end{eqnarray*}}

\begin{document}

\title{ Non-local kinetic energy functional from the Jellium-with-gap model: applications to 
Orbital-Free Density Functional Theory}
\author{Lucian A. Constantin}
\affiliation{Center for Biomolecular Nanotechnologies @UNILE, Istituto Italiano di Tecnologia, Via 
Barsanti, I-73010 Arnesano, Italy}
\author{Eduardo Fabiano}
\affiliation{Institute for Microelectronics and Microsystems (CNR-IMM), Via Monteroni, Campus Unisalento, 73100 Lecce, Italy.}
\affiliation{Center for Biomolecular Nanotechnologies @UNILE, Istituto Italiano di Tecnologia, Via Barsanti, I-73010
Arnesano, Italy}
\author{Fabio Della Sala}
\affiliation{Institute for Microelectronics and Microsystems (CNR-IMM), Via Monteroni, Campus Unisalento, 73100 Lecce, Italy.}
\affiliation{Center for Biomolecular Nanotechnologies @UNILE, Istituto Italiano di Tecnologia, Via Barsanti, I-73010
Arnesano, Italy}

\date{\today}

\begin{abstract}
Orbital-Free Density Functional Theory (OF-DFT) promises to describe the electronic structure of very 
large quantum systems, being its computational
cost linear with the system size. However, the OF-DFT accuracy strongly depends on the approximation 
made for 
the kinetic energy (KE) functional. To date, the most accurate KE functionals  are non-local functionals 
based on the linear-response kernel of the homogeneous electron gas, i.e. the jellium model.
Here, we use the linear-response kernel of the jellium-with-gap model, to  construct a simple non-local 
KE functional (named KGAP) which depends on the band gap energy.
In the limit of vanishing energy-gap (i.e. in the case of metals), the KGAP is equivalent to the 
Smargiassi-Madden (SM) functional, which is accurate for metals.
For a series of semiconductors (with different energy-gaps), the KGAP performs much better than SM, and 
results are close to the state-of-the-art functionals with complicated density-dependent kernels. 
%
%
\end{abstract}

\pacs{71.10.Ca,71.15.Mb,71.45.Gm}

\maketitle

\section{Introduction}

The main quantity in the Density Functional Theory (DFT) 
\cite{hohenberg1964inhomogeneous,levy1979universal} is the
ground-state electron density $n(\R)$.
In the most straightforward realization of DFT, i.e. orbital-free (OF) DFT,
the electron density is found by solving the Euler equation 
\cite{levy1979universal}
\begin{equation}
\frac{\delta T_s[n]}{\delta n(\R)}+v_{ext}(\R)+\int d\R' \frac{n(\R')}{|\R-\R'|}+\frac{\delta 
E_{xc}[n]}{\delta
n(\R)}=\mu,
\label{eq1}
\end{equation}
where $T_s[n]$ is the non-interacting kinetic energy (KE) functional, $v_{ext}(\R)$ is the external 
potential,
$E_{xc}[n]$ is the exchange-correlation (XC) energy functional, and $\mu$ is a Lagrange multiplier 
fixed from the
normalization condition $\int d\R n(\R)=N$, with $N$ being the total number of electrons. 
Both $T_s[n]$ and $E_{xc}[n]$ are unknown and they must be approximated. Many
valuable approximations, even at the semilocal level of theory, 
have been developed for $E_{xc}$ \cite{scuseriaREVIEW05,della2016kinetic}.
On the other hand, accurate approximations of $T_s[n]$ are much harder to
obtain \cite{wang2002orbital,karasiev2014progress,snyder2012finding,yao2016kinetic}, 
because this term usually gives the dominant contribution to the
ground-state energy \cite{levy1979universal} and especially because of its highly non-local nature
\cite{garcia1996kinetic,wang2002orbital,howard2001r,march1991non,snyder2012finding,della2015kohn}.

This problem is bypassed in the Kohn-Sham (KS) \cite{kohn1965self} formalism
where the non-interacting KE is treated exactly via the one-particle orbitals of
an auxiliary system of non-interacting particles. In this way, practical DFT
calculations in both quantum chemistry and material science are made
routinely possible \cite{tran2016rungs,burke2012perspective}. However, one has to pay the
cost of the introduction of orbitals into the formalism, which makes the method
formally scale as $\mathcal{O}(N^3)$. To overcome this limit, various  
fast electronic structure approaches have been developed, 
such as linear scaling $\mathcal{O}(N)$ methods based on density matrix 
approximations \cite{goedecker1999linear,goedecker2003linear}, as well as 
tight-binding and semi-empirical methods 
\cite{vogl1983semi,murrell1998semi,wahiduzzaman2013dftb}. 
Anyway, these
methods are numerically quite cumbersome if compared to the elegant OF-DFT approach.
Thus, intensive investigations are performed in the field of 
KE functional approximations suitable to  OF-DFT \cite{wang2002orbital,karasiev2014progress}, and 
important OF-DFT large-scale applications have been studied   
\cite{hung2009accurate,lambert2006structural,chen2013melting,gavini2007quasi,radhakrishnan2010effect,
gavini2008role,radhakrishnan2016orbital,caspersen2005finding,xiang2016understanding}. 

Kinetic energy functionals can be written in the general form
\begin{equation}
T_s[n] = \int\tau[n](\R)d\R \ ,
\end{equation}
where $\tau$ is the KE density, which is formally defined as $\tau(\R)=\sum_i |\nabla 
\phi_i(\R)|^2/2$, with $\phi_i(\R)$ being the $i$-th occupied Kohn-Sham orbital. For other (equivalent) 
formal definitions of $\tau$, see for example Refs. \cite{ayers2002local,anderson2010ambiguous}.
 
Approximations of $\tau(\R)$ 
can be classified on a ladder of
complexity. The first rung contains functionals whose KE density is only a 
function of the density $\tau(\R)=\tau(n(\R))$, such as Thomas-Fermi (TF)  
\cite{thomas1927calculation,fermi1927metodo}.
The TF theory \cite{thomas1927calculation,fermi1927metodo,lieb1973thomas} can not bind 
atoms into
molecules \cite{teller1962stability}, although it is asymptotically correct for heavy atoms and molecules 
\cite{lieb1976stability,lieb1981thomas,lieb1977thomas,heilmann1995electron,constantin2010communication} 
and accounts for the stability of bulk matter \cite{lieb1976stability}. Nevertheless, for the simple 
extension of the TF theory with the von Weizs\"{a}cker kinetic energy \cite{weizsacker1935theorie}, 
Lieb et al. have  
mathematically proven the existence of binding for two very dissimilar atoms 
\cite{benguria1981thomas}. This fact encouraged  
the investigation of exact KE properties \cite{levy1988exact,herring1986explicit,levy1985hellmann,
fabianoPRA13,ayers2005generalized,ayers2002local,anderson2010ambiguous,levy2001pair,
nagy1992relation,nagy2010pauli,nagy2015fisher}, and the development of semilocal KE functional approximations.
The simplest of them are found on the second rung of the ladder and are mainly
represented by the generalized gradient approximations (GGAs), of the form
$\tau^{GGA}(\R)=\tau(n(\R),\nabla n(\R))$. Starting with von Weizs\"{a}cker 
\cite{weizsacker1935theorie} and second-order gradient expansion \cite{Kirz57}, there are many GGA 
functionals constructed from exact conditions 
\cite{constantinPRL11,laricchia2011generalized,ou1991approximate,perdew1992generalized,ernzerhof2000role}, 
model systems \cite{constantin2009kinetic,vitos2000local,lindmaa2014quantum,PRBLGAP}, and empirical 
considerations \cite{borgooJCTC13,thakkar1992comparison,LC94,seino2018semi}. 
Often these semilocal functionals display several drawbacks and have limited applicability
in the context OF-DFT calculations \cite{xia2015single}.
However, some notable exceptions also exist \cite{xia2015single,trickey2015comment,xia2015reply,
karasiev2014progress,karasiev2014finite,karasiev2015chapter,karasiev2006born,karasiev2013nonempirical}. 
Among them we mention the VT84F GGA
of Ref. \cite{karasiev2013nonempirical}, and the vWGTF1 and vWGTF2 of Ref. \cite{xia2015single}, that can 
be considered state-of-the-art semilocal functionals for OF-DFT 
\cite{karasiev2014progress,karasiev2014finite,xia2015single,xia2015reply}.
On the third rung of the ladder are the Laplacian-level meta-GGA functionals, with the form 
$\tau^{MGGA}(\R)=\tau(n(\R),\nabla n(\R),\nabla^2 n(\R))$. The most known meta-GGA is the 
fourth-order gradient expansion of the uniform electron gas \cite{brack76,Ho}, that 
had been applied to metallic clusters in the OF-DFT context \cite{engel1991theory,engel1994gradient}. 
Several meta-GGAs 
have been recently developed 
\cite{cancio2016visualization,cancio2017visualisation,laricchia2013laplacian,constantinPRL17} for 
various purposes, including OF-DFT for solids \onlinecite{constantinPRL17}.
The next rung includes the class of u-meta-GGA functionals. Such approximations have been 
recently proposed \cite{constantin2016hartree,constantin2017modified} and they use as additional ingredient the 
Hartree 
potential $u(\R)=\int d\R' n(\R')/|\R-\R'|$, such that the KE density has the form
\begin{equation}
\tau^{uMGGA}(\R)=\tau(n(\R),\nabla n(\R),\nabla^2 n(\R),u(\R))\ .
\end{equation}
The u-meta-GGAs are promising tools for OF-DFT, but they  
require further investigations before they can become practical tools for
these calculations.  

Up to this level, the ladder of KE functionals contains only semilocal
approximations, i.e. functions that use as input ingredients only the electron density at
a given point in the space and other quantities (typically spatial
derivatives of the density) computed at the same point.
These approximations are computationally very advantageous because of their
local nature, and are theoretically justified by the the concept of nearsightedness of electrons, which 
means that the density $n(\R)$ depends significantly only on the effective external potential at nearby 
points \cite{prodan2005nearsightedness}. Consequently, any local physical property at point $\R$ can be 
described by the density behavior in a small volume $dV$ around this point. 
However, this principle does not hold in general and, especially for KE \cite{della2015kohn,constantin2016kinetic}, the 
non-local effects can not be ruled out in all cases. The consequence is that
semilocal KE functionals face several limitations.
In fact, in view of overcoming this problem, the u-meta-GGA
already contains important non-locality through the Hartree-potential ingredient. 
The fundamental solution, anyway, is to consider fully non-local KE
approximations.

Nowadays, the most sophysticated KE functionals are the fully non-local ones 
\cite{huang2010nonlocal,wang1998orbital,shin2014enhanced,ho2008analytic,alonso1978nonlocal,
garcia1996nonlocal,chacon1985nonlocal,garcia1998nonlocal,garcia2008approach,genova2017nonlocal,
ludena2017kinetic,salazar2016study}. Most of them can be written in the generic form  
\begin{equation}
T_s[n]=T_s^{TF}+T_s^W+\langle n(\R)^\alpha|w(\R-\R',n(\R),n(\R'))|n(\R')^\beta\rangle,
\label{eq2}
\end{equation}
where $T_s^{TF}=\frac{3}{10}(3\pi^2)^{2/3}\langle n(\R)^{5/3}\rangle$, and $T_s^W=\langle\frac{|\nabla 
n(\R)|^2}{8n(\R)}\rangle$ 
are the TF and von Weizs\"{a}cker functionals respectively, $\alpha$ and $\beta$ are parameters, and 
the kernel $w(\R-\R',n(\R),n(\R'))$ is chosen such that $T_s[n]$ recovers the exact linear response (LR) of 
the non-interacting uniform electron gas without exchange 
\cite{wang1998orbital,wang1999orbital,huang2010nonlocal} 
\begin{equation}
\hat{\mathcal{F}} \left( \frac{\delta^2 T_s[n]}{\delta n(\R)\delta
n(\R')}|_{n_0}\right)=-\frac{1}{\chi_{Lind}}=\frac{\pi^2}{k_F}F^{Lind}(\eta),
\label{eq3}
\end{equation}
with
\begin{equation}
F^{Lind}=\left( \frac{1}{2}+\frac{1-\eta^2}{4\eta}\ln \left| \frac{1+\eta}{1-\eta} \right| 
\right)^{-1}
\label{eq4}
\end{equation}
being the Lindhard function \cite{lindhard1954properties,wang2002orbital},
$\eta=q/(2k_F)$ being the 
dimensionless momentum ($q$ is  the momentum and $k_F=(3\pi^2 n_0)^{1/3}$ is the Fermi
wave vector of the jellium with the constant density $n_0$), and $\hat{\mathcal{F}}(\cdot)$ denotes
the Fourier transform. 
The most simple functionals having the form of Eq. (\ref{eq2}) are the ones with 
a density-independent kernel $w(\R-\R')$, which are also the most attractive
from the computational point of view. 
After using the constraint of Eq. (\ref{eq3}), they depend only on the choice of the parameters $\alpha$ and
$\beta$. 
The most known functionals of this class are:
\begin{itemize}
\item{The Perrot functional \cite{perrot1994hydrogen}, with $\alpha=\beta=1$;}
\item{The Wang-Teter (WT) functional \cite{wang1992kinetic}, with $\alpha=\beta=5/6$;}
\item{The Smargiassi and Madden (SM) functional \cite{smargiassi1994orbital}, with $\alpha=\beta=1/2$;}
\item{The Wang-Govind-Carter (WGC) functional \cite{wang1999orbital}, with
  $\alpha,\beta=5/6 \pm \sqrt{5}/6$.}
\end{itemize}
In the context of orbital-free DFT, these KE functionals are usually accurate for structural 
properties of simple metals \cite{wang1999orbital}, systems for which the LR of jellium is an 
excellent model. However, they may fail for other bulk solids, such as
semiconductors and insulators, where the 
jellium perturbed by a small-amplitude, short-wavelength density wave is not a relevant model. 

To improve the description of semiconductors, the Huang-Carter (HC) functional has been introduced 
\cite{huang2010nonlocal}. The kernel of this functional is more complicated, being dependent on the 
density and the gradient of the density, as well as on empirical parameters. 
This functional, as any non-local KE functional \cite{wang1999orbital}, has a quasi-linear scaling 
with system size ($N$),
behaving as $\mathcal{O}(N\ln(N))$, but its prefactor may be quite large \cite{shin2014enhanced}, 
lowering considerably the overall computational efficiency.
%
%
Consequently, it is 
significantly slower than non-local KE functionals with density-independent kernels.

Computationally efficient methods/functionals for semiconductors have been recently 
developed. Thus, the density-decomposed WGCD KE functional \cite{xia2012density}, as well as the enhanced 
von 
Weizs\"{a}cker-WGC (EvW-WGC) KE functional \cite{shin2014enhanced} are both based on the WGC 
density-dependent kernel \cite{wang1999orbital}. These 
functionals, which also contain several empirical parameters, are hundred times faster than HC, providing 
similar accuracy as the HC functional, for semiconductors. Additionally the EvW-WGC functional accurately 
describes metal-insulator transitions \cite{shin2013mechanical,shin2014enhanced}. However, we mention 
that, in contrast to 
HC, the WGCD and EvW-WGC can not be written in the form of Eq. (\ref{eq2}).     

In this article we introduce a non-local KE functional with a
density-independent kernel (KGAP) that 
recovers not Eq. (\ref{eq3}) but the LR of the jellium-with-gap model
\cite{levine1982new,PRBLGAP}. 
This is an important generalization of the uniform electron gas, that has already been  used
to have qualitative and quantitative insight for semiconductors
\cite{callaway1959correlation,penn1962wave,srinivasan1969microscopic,levine1982new,tsolakidis2004effect},
to develop an XC kernel for the optical properties of materials
\cite{trevisanutto2013optical}, and to construct accurate functionals for the
ground-state DFT
\cite{rey1998virtual,krieger1999electron,krieger2001density,toulouse2002validation,toulouse2002new,
fabiano2014generalized}. 
Recently, the KE gradient expansion of the jellium-with-gap has also been
derived and used in the construction of semilocal KE functionals
\cite{PRBLGAP}.
The KGAP functional fulfills important exact properties and shows a better
accuracy as well as a broader applicability than other existing non-local
functionals with a density independent kernel. 

The paper is organized as follows. In Section \ref{sect_theory}, we construct the KGAP functional, and 
in Section \ref{sect_results} we test it for equilibrium lattice constants and bulk moduli of several 
bulk solids, performing full OF-DFT calculations. Computational details of these calculations 
are presented in Section \ref{sect_compdet}. Finally, in Section \ref{sect_conclusions} we 
summarize our results.

\section{Theory}
\label{sect_theory}
Let us consider a generalization of Eq. (\ref{eq2}) of the form
\begin{equation}
T_s[n]=\lambda T_s^{TF}+\mu T_s^W+\langle n(\R)^\alpha|w(\R-\R',E_g)|n(\R')^\beta\rangle \ ,
\label{eq5}
\end{equation}
where $\lambda,\mu \in [0,1]$ as well as $\alpha$ and $\beta$ 
are positive parameters, and $w(\R-\R',E_g)$ is a density-independent kernel 
chosen such that the whole KE functional $T_s[n]$ satisfies the LR of the jellium-with-gap model 
\cite{PRBLGAP,levine1982new}
%
\begin{equation}
\hat{\mathcal{F}} \left( \frac{\delta^2 T_s[n]}{\delta n(\R)\delta 
n(\R')}|_{n_0}\right)=-\frac{1}{\chi_{GAP}}=\frac{\pi^2}{k_F}F^{GAP}(\eta),
\label{eq6}
\end{equation}
with 
\begin{eqnarray}
1/F^{GAP} & = & \frac{1}{2}-\frac{\Delta(\arctan(\frac{4\eta+4\eta^2}{\Delta})+
\arctan(\frac{4\eta-4\eta^2}{\Delta}))}{8\eta}\nonumber+ \\
&& + (\frac{\Delta^2}{128\eta^3}+\frac{1}{8\eta}-\frac{\eta}{8})\ln(\frac{\Delta^2+(4\eta+4\eta^2)^2}
{\Delta^2+(4\eta-4\eta^2)^2}),
\label{eq7}
\end{eqnarray}
where $\Delta=2E_g/k_F^2$, with $E_g$ being the gap.
In momentum space, the kernel $w$ is
\begin{eqnarray}
w(\Q)&=&-\frac{\chi_{GAP}^{-1}-\lambda\chi_{TF}^{-1}-\mu\chi_{W}^{-1}}{2\alpha\beta n_0^{\alpha+\beta-2}}= 
\nonumber\\
& = & \frac{5}{9\alpha\beta n_0^{\alpha+\beta-5/3}}(F^{GAP}(\eta)-\lambda-3\mu\eta^2),
\label{eq8}
\end{eqnarray}
with $\chi_{TF}=-k_F/\pi^2$, and $\chi_{W}=-k_F/(3\pi^2\eta^2)$. Some details
on the derivation of Eq. (\ref{eq8}) are given in Appendix A.

A careful analysis of $F^{GAP}$ is provided in
Ref. \onlinecite{PRBLGAP}. 
The most important features  of $F^{GAP}$ are also summarized in the Appendix B. 
Here we use them, together with the procedure proposed in 
Refs. \onlinecite{wang1998orbital,wang1999orbital}, to find the low-$q$ (at $\Delta\rightarrow 0$) 
and high-$q$ (at any $\Delta$) limits of the functional of Eq. (\ref{eq5}).
Some details of the derivation of these limits are given in Appendix C.
The mentioned limits are:
\begin{widetext}
\begin{eqnarray}
\nonumber
\lim_{\Q\rightarrow 0}T_s[n]& \rightarrow & \big[\lambda+\frac{5}{9\alpha\beta}(1-\lambda)\big]T_s^{TF}+\frac{T_s^W}{9}+  
\frac{5(1-\lambda)}{9\alpha\beta} (\alpha+\beta-\frac{5}{3})\big\{ <\delta n|\tau_{TF}>+
(\alpha+\beta-\frac{8}{3})\frac{<\delta^2 n|\tau_{TF}>}{2} \big\}+ \\
\label{eq9}
&& +  (\frac{1}{9}-\mu)(\alpha+\beta-1)\big \{ <\delta n|\tau_W> + (\alpha+\beta-2)\frac{<\delta^2 
n|\tau_W>}{2}\big\} + \mathcal{O}(\delta^3 n)\ ,\\
\nonumber
\lim_{\Q\rightarrow \infty}T_s[n] & \rightarrow & T_s^W+\big(\lambda-\frac{1}{3\alpha\beta}-\frac{5\lambda}{9\alpha\beta}\big)T_s^{TF}-\frac{3+5\lambda}{9\alpha\beta}(\alpha+\beta-\frac{5}{3})\big\{ <\delta n|\tau_{TF}> + (\alpha+\beta-\frac{8}{3})\frac{<\delta^2 n|\tau_{TF}>}{2} \big\}+\\
\label{eq11}
&& +(1-\mu)(\alpha+\beta-1) \big\{ <\delta n|\tau_W>+(\alpha+\beta-2)\frac{<\delta^2 n|\tau_W>}{2}   \big\}+ \mathcal{O}(\delta^3 n)\ ,
\end{eqnarray}
\end{widetext}
where $\delta n=n(\R)/n_0-1$. 
Equation (\ref{eq9}) is a generalization of Eq. (17) of Ref. \onlinecite{wang1999orbital}, 
recovering it for the case $\lambda=\mu=1$. In this low-$q$ limit the exact behavior is described by 
the second-order gradient expansion (GE2) (i.e. $T_s^{GE2}=T_s^{TF}+T_s^W/9$). This is recovered whenever
\begin{equation}
\lambda=1 ,\;\;\;\rm{and}\;\;\; (\frac{1}{9}-\mu)(\alpha+\beta-1)=0\ ,
\label{eq10}
\end{equation}
or, independently on the values of $\alpha$ and $\beta$, when $\lambda=1$ and $\mu=1/9$.
The only functional with the form of Eq. (\ref{eq5}) that is correct in the
limit $\Q\rightarrow 0$ is the SM functional ($\alpha=\beta=1/2$, $\lambda=1$,
$\mu=1$, and $E_g=0$).
Note, anyway, that for $\Q\rightarrow0$ we have $T_s^{TF}\gg T_s^{W}$, thus any functional recovering correctly
the TF behavior is accurate.

For the case $\Q\rightarrow\infty$, $F^{GAP}$ behaves as $F^{Lind}$ for any $\Delta$, and we 
recover Eq. (18) of Ref. \onlinecite{wang1999orbital} when $\lambda=\mu=1$.
The exact LR behavior \cite{PRBLGAP} is
\begin{equation}
\label{e14}
\lim_{\Q\rightarrow \infty}T_s[n] \rightarrow T_s^W-\frac{3}{5}T_s^{TF}\ ,
\end{equation}
which can be satisfied if
\begin{eqnarray}
&& \lambda-\frac{3+5\lambda}{9\alpha\beta}=-\frac{3}{5}, 
\nonumber\\
&& (\alpha+\beta-\frac{5}{3})\frac{3+5\lambda}{9\alpha\beta}=0,
\nonumber\\
&& (1-\mu)(\alpha+\beta-1)=0. 
\label{eq12}
\end{eqnarray}
Only the WGC functional ($\alpha,\beta=5/6 \pm \sqrt{5}/6$, $\lambda=\mu=1$, $E_g=0$) is correct in the 
limit $\Q\rightarrow \infty$. We also remark that, in this limit, $T_s^W\gg T_s^{TF}$, therefore, 
in principle, any functional with the form of Eq. (\ref{eq5}) and with $\mu=1$ does not fail badly in this limit. 
 
Inspection of Eqs. (\ref {eq10}) and (\ref {eq12}) shows that it is not possible to fix the parameters
in order to satisfy exactly both the low- and high-$q$ limits. 
Nevertheless, the choice 
\begin{equation}
\label{eell}
\lambda=1 \;\;\; , \;\;\; \mu=1
\end{equation}
allows to recover in both cases the correct leading term, guaranteeing that $T_s[n]$ performs reasonably 
well in 
both limits, independently on $\alpha$ and $\beta$ ($\alpha,\beta> 0$). 
This choice appears then to be the most physical for a kinetic functional.
Moreover, in the low-$q$ limit the correct behavior is anyway obtained fixing $\alpha=\alpha^{LQ}=1/2$ and
$\beta=\beta^{LQ}=1/2$; similarly, in the high-$q$ limit this occurs for
$\alpha=\alpha^{HQ}=5/6+\sqrt{5}/6$ and $\beta=\beta^{HQ}=5/6-\sqrt{5}/6$.

We can use these observations to propose a new kinetic functional based on Eq. (\ref{eq5}).
This is named KGAP and uses $\lambda=1$ and $\mu=1$ as well as 
\begin{eqnarray}
\label{eeb}
\alpha^{KGAP} & = & \alpha^{LQ}+(\alpha^{HQ}-\alpha^{LQ})\frac{E_g^2}{b+E_g^2} \\
\beta^{KGAP} & = & \beta^{LQ}+(\beta^{HQ}-\beta^{LQ})\frac{E_g^2}{b+E_g^2} \ ,
\label{eea}
\end{eqnarray}
where $b=5$ eV$^2$, is a parameter that controls the connection between the low- and high-$q$ limits.
Overall the KGAP functional satisfies the following conditions:
(1) for metals ($E_g=0$), $F^{GAP}=F^{Lind}$ and  KGAP performs as the SM functional, recovering 
GE2 for slowly-varying densities; (2) for semiconductors and insulators, KGAP is correct at $\Q\rightarrow 0$ 
(see Eqs. (B-1) and (B-2)). This important exact condition is very difficult to be fulfilled, and even the 
HC functional constructed for semiconductors can not satisfy it \cite{huang2010nonlocal};
(3) for large-gap insulators, KGAP correctly recovers the 
exact behavior of Eq. (\ref{e14}).

\section{Computational details}
\label{sect_compdet}
The KGAP functional has been implemented in PROFESS 3.0
(PRinceton Orbital-Free Electronic Structure Software), a plane-wave-based OF-DFT code 
\cite{ho2008introducing}.
We have then tested it for the simulation of cubic-diamond Si, various III-V cubic zincblende
semiconductors (AlP, AlAs, AlSb, GaP, GaAs, GaSb, InP, InAs, and InAs) \cite{shin2014enhanced}, and 
several metals,  namely Al, Mg and Li, in their
simple-cubic (sc), face-centered-cubic (fcc), and body-centered-cubic (bcc) structures.
The results have been compared to those obtained with the SM and HC functionals.
In this work, we use the HC with optimized parameters for semiconductors
\cite{huang2010nonlocal} ($\lambda=0.01177$, and $\beta=0.7143$).
On the other hand, using the Perrot, WT, or WGC functionals almost no
well converged result could be obtained for the tested semiconductors.

For a better comparison with literature results, we have used in all calculations
the Perdew and Zunger XC LDA parametrization \cite{perdew1981self}, 
bulk-derived local pseudopotentials (BLPSs), as in Refs.
\cite{xia2015single,shin2014enhanced} and plane wave basis kinetic energy cutoffs of 1600 eV. 
Equilibrium volumes and bulk moduli have been calculated by expanding and compressing the optimized 
lattice parameters by up to about 10\% to obtain thirty energy-volume points and then fitting with the 
Murnaghan’s equation of state \cite{murnaghan1944compressibility}.

\section{Results}
\label{sect_results}

\subsection{Energy gap}
%
\begin{figure}
\includegraphics[width=\columnwidth]{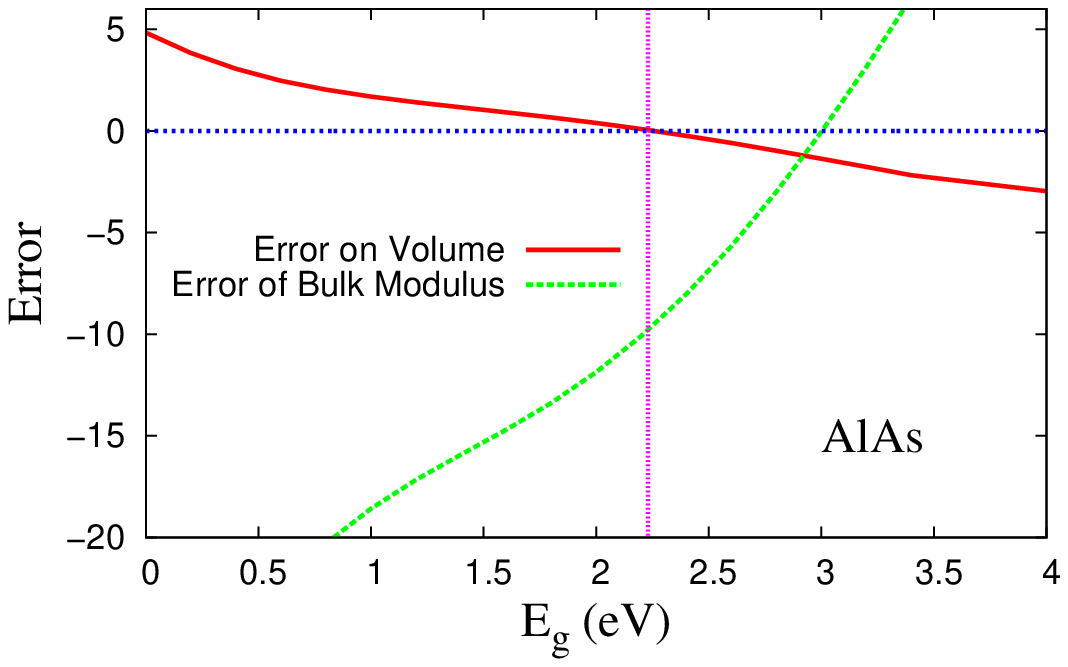}
\includegraphics[width=\columnwidth]{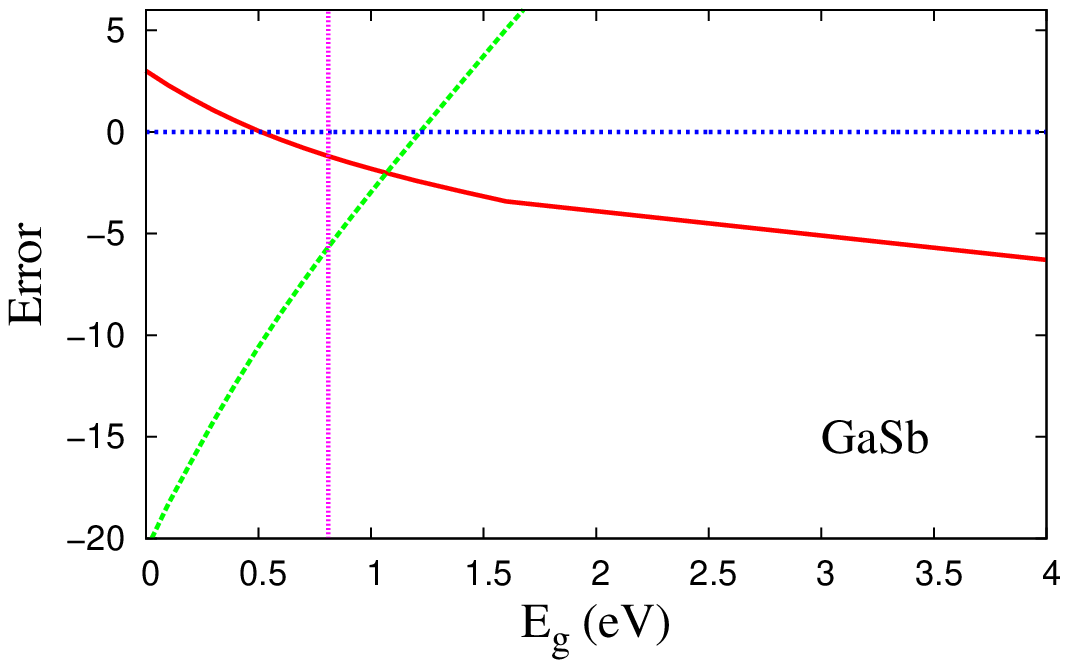}
\caption{Errors of KGAP OF-DFT calculations with respect to the KS-DFT references (OFDFT-KSDFT) for equilibrium
volumes ( \AA$^3$ /cell, red lines), and for bulk modulus (GPa, green lines), as a function of the energy gap
parameter $E_g$ (in eV), for the AlAs and GaSb semiconductors. The experimental fundamental bad gaps are shown
with horizontal lines ($E_g=2.23$ eV and 0.81 eV for AlAs and GaSb, respectively).
}
\label{fg}
\end{figure}
%
The KGAP functional, defined by Eqs. (\ref{eq5}), (\ref{eq8}), (\ref{eell})-(\ref{eea}), 
depends on the energy-gap $E_g$. 
Previous investigation on exchange-correlation kernel
indicated that  $E_g$ can be fixed to the  experimental 
fundamental gap of semiconductors and insulators \cite{trevisanutto2013optical}.
In this subsection we will verify if this can be considered a good approximation also for the kinetic 
energy. 

In Fig. \ref{fg} we report the errors on equilibrium volumes (\AA$^3$/cell) and bulk moduli (GPa) 
for AlAs and GaSb, as a function
of the parameter $E_g$.
We recall that setting $E_g=0$ the KGAP functional is equivalent to the SM functional, which is not 
accurate for semiconductors, as shown in  Fig. \ref{fg}.
When $E_g$ is increased the errors descrease for both systems 
and properties vanishing near the the vertical lines, which indicate the 
experimental fundamental gaps of AlAs (2.23 eV) and GaSb (0.81eV). 
Similar results are obtained for other semiconductors.

%
\begin{figure}
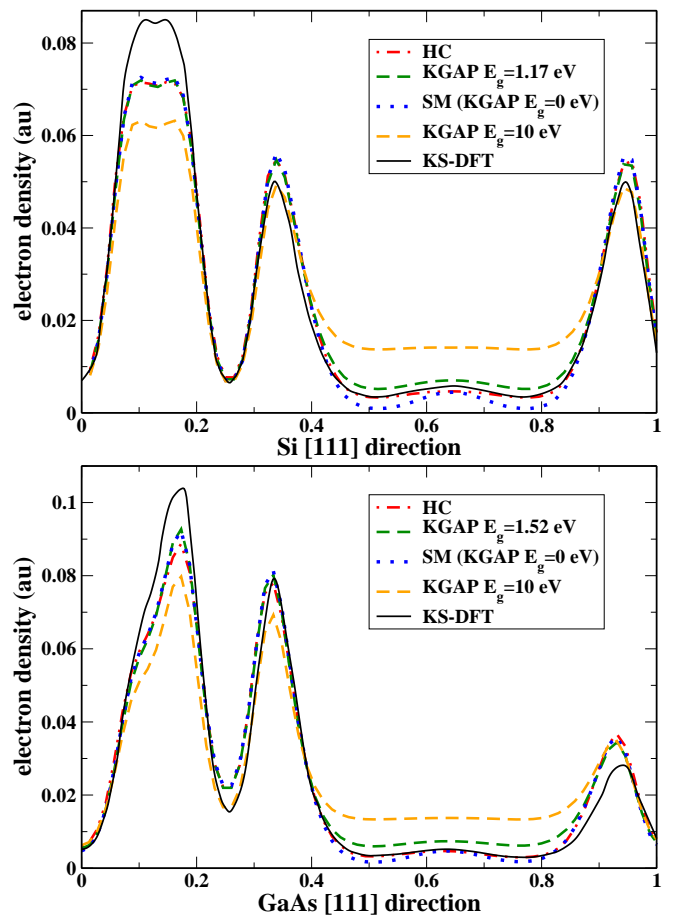

\includegraphics[width=\columnwidth]{si_fig.eps}
\includegraphics[width=\columnwidth]{fig_gaas.eps}
\caption{Electron densities of Si (upper panel) and GaAs (lower panel) along 
the [111] direction, obtained from OF-DFT calculations with several KE functionals. 
The results for KGAP use the exact experimental band gaps ($E_g=1.17$ eV for 
Si, and 1.52 eV for GaAs, respectively), the vanishing band gap case ($E_g=0$) which represents the 
SM functional, and the case $E_g=10$ eV.
For comparison, see also Figs. 8 and 9 of Ref. \cite{shin2014enhanced}.
}
\label{figd}
\end{figure}
%
Next, in Fig. \ref{figd}, we show the OF-DFT densities of Si and GaAs along
the [111] direction, computed with several KE functionals. In both panels, all functionals with the exception of 
the $E_g=10$ eV extreme case, agree well in most of the space and more significant differences are obtained only 
at the bonding region, in the range between 0.4 and 0.8. Here the SM functional (i.e. the KGAP with 
$E_g=0$) gives 
smaller densities than the HC ones, with pronounced oscillatory features. On the other hand, the KGAP 
functional with the exact experimental band gap, gives accurate densities, being of comparable accuracy as 
the HC ones. We also note that the results obtained from KGAP with $E_g=10$ eV are 
inaccurate because of an unrealistic value of the $E_g$ parameter.
Nevertheless, even in this extreme case, the densities are smooth and the calculations are numerically 
stable. These facts are strong indications that $F^{GAP}$ 
is an useful, well-behaved generalization of $F^{Lind}$.

The results of Figs. \ref{fg} and \ref{figd} show that the experimental fundamental gap is a good choice 
for the  $E_g$ parameter of the KGAP functional. Hence, unless 
differently stated, in all our calculations we fixed $E_g$ to the 
experimental fundamental gap value of the investigated material.    
Finally we mention that, due to its $E_g$ dependence, the KGAP should be seen as a 
semi-empirical functional.

\subsection{Global Assessment for Semiconductors and Metals}
%
\begin{figure}
\includegraphics[width=\columnwidth]{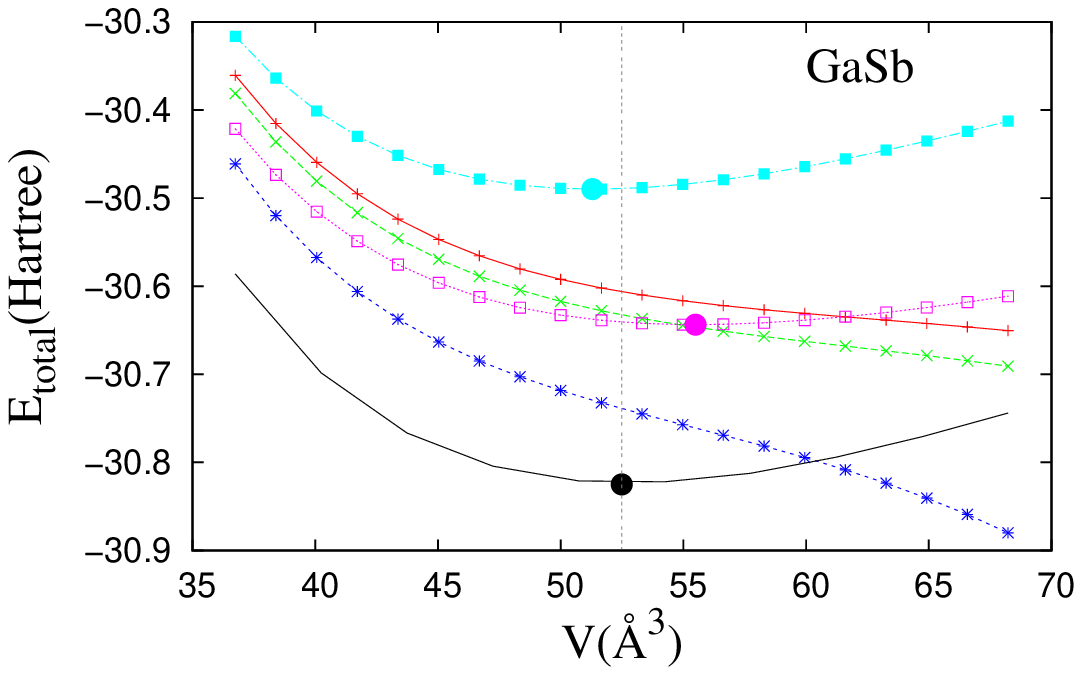}
\includegraphics[width=\columnwidth]{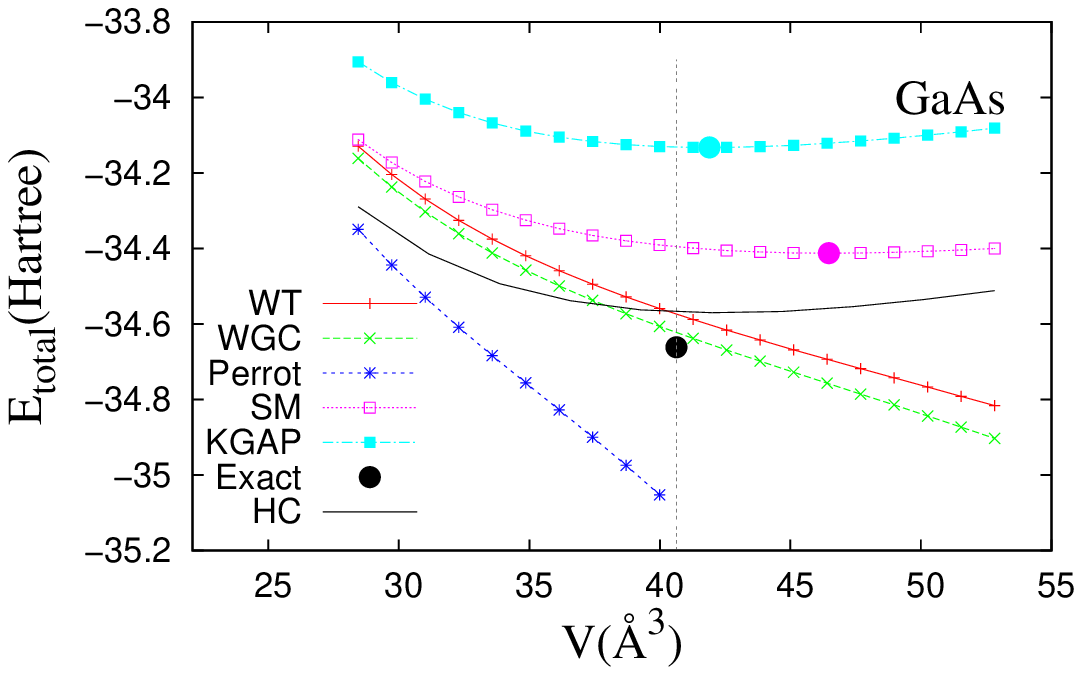}
\includegraphics[width=\columnwidth]{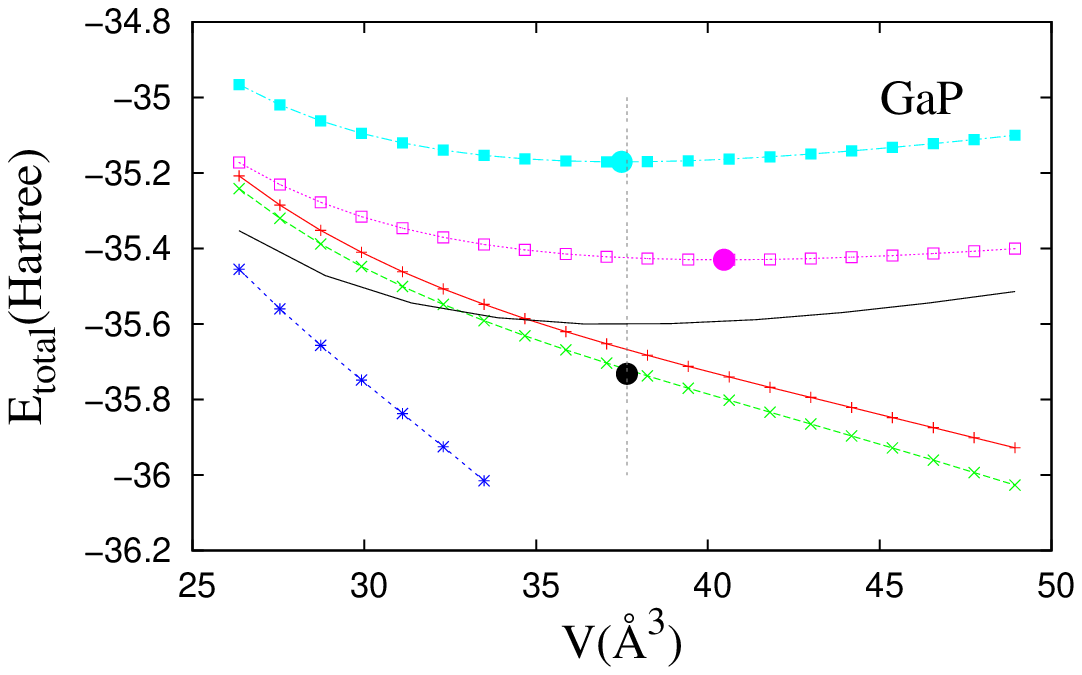}
\caption{Total energy (in Hartree) versus the volume of the unit cell (in \AA$^3$) computed using OF-DFT 
calculations with several non-local functionals with density-independent kernels (Perrot 
\cite{perrot1994hydrogen}, WT \cite{wang1992kinetic},
WGC \cite{wang1999orbital}, SM \cite{smargiassi1994orbital}, and KGAP) for 
GaSb (upper panel), 
GaAs (middle panel), and GaP (lower panel). 
The KS-DFT equilibrium point (denoted as Exact) is shown with black big-dot. 
The SM and KGAP equilibrium points are also emphasized with big-dots. 
For comparison, we also show the results of the HC \cite{huang2010nonlocal} state-of-the-art non-local KE 
functional with a density-dependent kernel. 
}
\label{f1}
\end{figure}
%
In Fig. \ref{f1} we show the total 
energy versus volume curves for GaSb, GaAs and GaP bulk solids, computed using various KE functionals.
We observe that for all three cases, the Perrot, WT, and WGC functionals do not predict any binding. 
Moreover, their failures are accentuated when the fundamental band gap of the material $E_g$ increases. 
For example, the Perrot functional gives converged results for GaSb 
($E_b=0.81$ eV), while it converges only within few points in the cases of GaAs ($E_b=1.52$ eV) 
and GaP ($E_b=2.35$ eV). Also, the quality of the WT and WGC results diminishes for GaAs and GaP in 
comparison with GaSb. 
On the other hand, SM always yields a bound result but the potential energy curves are generally rather 
flat and the minima are always moved towards too large volumes.
Finally, KGAP can consistently reproduce the reference values with good accuracy.
Nevertheless, inspection of the figures shows that the KGAP functional yields a quite systematic overestimation of 
the total energies, giving a shift about 0.5 Hartree towards higher energies. Such a behavior is explained by Eq. 
(B-6). This feature is anyway not a serious flaw for the functional, since absolute energies are rarely 
important, whereas relative energies (such as in potential energy curves) are well described by KGAP.

%
\begin{table}
\begin{center}
\caption{\label{tav}
Errors of OF-DFT with respect to the KS-DFT references ($\rm{OFDFT}-\rm{KSDFT}$) for equilibrium volumes 
(\AA$^3$/cell), computed from different KE functionals.
The KS reference values are reported in the last column, and the exact band gap energies (in eV) used for the 
KGAP functional are shown in the second column.
The last lines of every panel report the mean absolute errors (MAE).
}
\begin{ruledtabular}
\begin{tabular}{lrrrrr}
      &  $E_g$ (eV) & SM & KGAP & HC & KS \\ \hline
\multicolumn{5}{c}{Semiconductors}\\
Si   & 1.17  & 1.3  & -0.3 & 0.0  & 19.781 \\
GaP  & 2.35  & 2.8  & 0.9  &  0.8 & 37.646 \\
GaAs & 1.52  & 5.8  & 1.3  & -0.6 & 40.634 \\
GaSb & 0.81  & 3.0  & -1.0 &  0.7 & 52.488 \\
AlP  & 2.50  & 2.5  & -1.0 & 0.4  & 40.637 \\
AlAs & 2.23  & 4.8  & 0.2  & -1.1 & 43.616 \\
AlSb & 1.69  & 2.3  & -3.8 & 0.7  & 56.607 \\
InP  & 1.42  & 2.7  & 0.7  & 0.1  & 46.040 \\
InAs & 0.42  & 4.9  & 3.0  & -1.5 & 49.123 \\
InSb & 0.24  & 2.2  & 0.5  & 0.1  & 62.908 \\
MAE   &      & 3.24 & 1.27 &  0.63 & \\ 
\multicolumn{5}{c}{Metals}\\
Al-sc & 0 & 0.32 &  0.32 & -0.52 & 19.937 \\
Al-fcc & 0 & 1.95  & 1.95 & 2.44 & 16.575 \\
Al-bcc & 0 & 0.97 & 0.97 & 1.94 & 17.025 \\
Mg-sc & 0 & 0.62 & 0.62 & 1.07 & 27.107 \\
Mg-fcc & 0 & 1.28 & 1.28 & 1.28 & 23.073 \\
Mg-bcc & 0 & 1.42 & 1.42 & 1.25 & 22.939 \\
Li-sc & 0 & 0.20 & 0.20 & 0.46 & 19.932 \\
Li-fcc & 0 & 0.22 & 0.22 & 0.52 &  19.308 \\ 
Li-bcc & 0 & 0.22 & 0.22 & 0.51 & 19.397 \\
MAE    &   & 0.80 & 0.80	& 1.11 \\
\end{tabular}
\end{ruledtabular}
\end{center}
\end{table}
%
%
\begin{table}
\begin{center}
\caption{\label{tab}
Errors of OF-DFT with respect to the KS-DFT references ($\rm{OFDFT}-\rm{KSDFT}$) for  bulk moduli (GPa), 
computed from different KE functionals. 
The KS reference values are reported in the last column, and the exact band gap energies (in eV) used for the 
KGAP functional are shown in the second column.
The last lines of every panel report the mean absolute errors (MAE).
}
\begin{ruledtabular}
\begin{tabular}{lrrrrr}
      &  $E_g$ (eV) & SM & KGAP & HC & KS \\ \hline
\multicolumn{5}{c}{Semiconductors}\\
Si   & 1.17 & -42  & -14.2  & 0.9 & 98 \\
GaP  & 2.35 & -28  &  -2.8  & -14  & 80 \\
GaAs & 1.52 & -35  & -12.8 & -3   & 75 \\
GaSb & 0.81 & -21  & -6.4  & -6   & 56  \\
AlP  & 2.50 & -32  & -8.6  & 1    &  90 \\
AlAs & 2.23 & -33  & -10.8 & 4    & 80 \\
AlSb & 1.69 & -23  &  2.3  & -1    & 60 \\
InP  & 1.42 & -25  & -14.1 & 5    & 73 \\
InAs & 0.42 & -24  & -17.7 & 4    & 65 \\
InSb & 0.24 & -17  & -13.1 & 1    & 50 \\
MAE& & 27.91 & 10.28 &  4.00 & \\
\multicolumn{5}{c}{Metals}\\
Al-sc  & 0 &4.1 & 4.1  & 1.8 & 57 \\
Al-fcc & 0 &-13.8 & -13.8  & -28.0 & 77  \\
Al-bcc & 0 &-5.3 &  -5.3 & -24.4  & 70\\
Mg-sc & 0 &1.5 &  1.5  & 3.7 &  24\\
Mg-fcc &0 & -0.3 &  -0.3 &  -3.2 & 38\\
Mg-bcc &0 & 1.2 &  1.2 & -4.3 & 38\\
Li-sc & 0 &-0.1  &  -0.1  & -0.6 & 17\\ 
Li-fcc &0 &  0.2  & 0.2  & -0.2 & 17\\
Li-bcc &0 &  -0.5 &  -0.5 & -0.9  & 16\\
MAE  &       & 3.00 & 3.00  &  7.64       \\
\end{tabular}
\end{ruledtabular}
\end{center}
\end{table}
%
%
\begin{figure}
\includegraphics[width=\columnwidth]{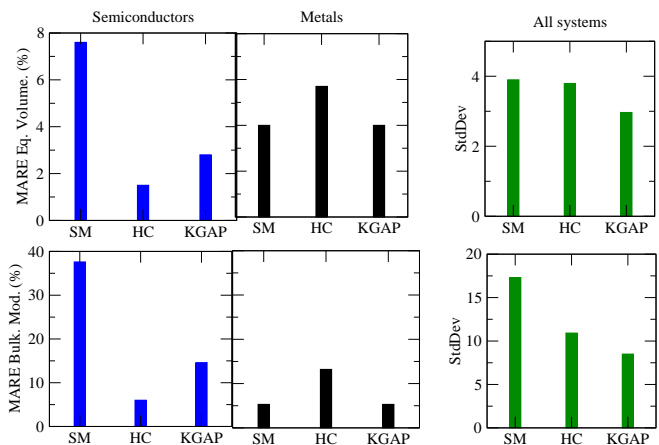}
\caption{Error statistics (mean absolute relative error (MARE) and standard deviation (StdDev)) of the OF-DFT 
calculations performed with 
SM \cite{smargiassi1994orbital}, HC \cite{huang2010nonlocal} and KGAP KE functionals. Full results are reported 
in Tables \ref{tav} and \ref{tab}. 
}
\label{fd}
\end{figure}
%
In Tables \ref{tav} and \ref{tab} we report the results  
for equilibrium volumes and bulk moduli
of semiconductors and simple metals.
The mean absolute relative errors (MARE) and the standard deviations (StdDev) are illustrated in Fig. \ref{fd}.

As shown in Fig. \ref{f1}, among the non-local KE functionals with a density-independent kernel constructed from 
the LR of the uniform electron gas, only the SM functional \cite{smargiassi1994orbital} shows converged results 
for semiconductors and a meaningful energy versus volume convex curve: for this reason this is the only one 
reported in this section. Anyway, the performance of the SM functional is quite modest for semiconductors, 
giving a MARE of 7.5\% for equilibrium volumes, and a MARE of 38.1\% for bulk moduli. 
On the other hand accurate results are obtained 
for metals with a MARE of 4.0\% for equilibrium volumes and MARE of 5.3\% for bulk moduli. Nevertheless, we 
recall that the WGC and WT KE functionals are in general better than the SM functional, for simple metals 
\cite{carling2003orbital}.

An opposite trend is obtained for the HC functional which has been developed for semiconductors 
\cite{huang2010nonlocal}.
The  MAREs for equilibrium volumes and bulk moduli are 1.4\% and 6\% for semiconductors, whereas
much bigger errors are found for metals (5.7\% and 13.2\%, respectively).
Thus, although the HC is very accurate for semiconductors, it is worse than SM for metals (improvement can be 
obtained employing dedicated fitting parameters).

The KGAP functional is significantly better than the SM functional for semiconductors. For equilibrium 
volumes the MARE is 2.7\%  and for bulk moduli the MARE is below 14.6\%, thus not far from the HC. 
By construction the KGAP functional is equivalent to SM for metals, so that KGAP is reasonably accurate for 
these systems. Note that for bulk moduli the mean absolute error is about 10 GPa, being comparable or even 
smaller than that due to the use of XC approximations in full KS-DFT calculations (see for example Table I of 
Ref. \cite{constantinPRB16}).

In the right panels of Fig.\ref{fd}, we report the standard deviations considering both semiconductors 
and metals, in order
to measure if a given functional describes different systems with similar accuracy.
The SM functional describes very differently metals and semiconductors, so the StdDev is large, in
particular for the bulk modulus (StdDev=17\%).
The HC functional has similar StdDev as SM functional for the equilibrium volume, whereas it is 
smaller for the bulk modulus (StdDev= 11\%).
On the other hand, the KGAP functional gives significantly reduced StdDev
for both properties.
\section{Conclusions}
\label{sect_conclusions}
We have constructed a simple non-local KE functional named KGAP, with a density-independent kernel found from 
the linear response of the jellium-with-gap model. This functional has the correct physics of metals, 
semiconductors and insulators in the $\Q\rightarrow 0$ limit, being also very accurate for small 
perturbations of the density with large wave-vectors. 
The KGAP functional performs well in the orbital-free DFT 
context, converging very fast and being equally accurate for metals (where by construction recovers the SM  
functional), and semiconductors. 
To our knowledge, the KGAP functional is the only one from the class of 
approximations with density-independent kernels, that has a rather broad applicability in solid-state physics. 

In this first implementation, the KGAP functional has been tested on simple bulk systems. 
In this case the KGAP semi-empirical functional requires the a priori knowledge of the $E_g$ 
parameter which can be well 
approximated by the fundamental band gap energy of the system. 
For more general applications (e.g. interfaces)  the $E_g$ parameter must be spatially dependent, as shown for 
example in Refs.\cite{PRBLGAP,fabiano2014generalized,terentjevs2014first}.
Such a KE functional, will be more complicated than the simple KGAP, but we expect it to be very accurate. 
We will address this important issue in next work.

  \renewcommand{\theequation}{A-\arabic{equation}}
  \setcounter{equation}{0}  
  \section*{APPENDIX A}  

Let consider a functional $J[n]$ of the form
\begin{equation}
J[n]=\int\int d\R d\R'\; n^\alpha(\R) w(\R-\R')n^\beta(\R'),
\label{aa1}
\end{equation}
with $\alpha$ and $\beta$ positive constants. Using the definition of 
functional derivative 
\begin{equation}
\int \;\frac{\delta J}{\delta n(\R)}\phi(\R)d\R=\frac{d}{d\epsilon}J[n+\epsilon\phi]\mid_{\epsilon=0},
\label{aa2}
\end{equation}
we find
\begin{equation}
\frac{\delta J}{\delta n(\R)}=\int d\R'\;w(\R-\R')\{\alpha n(\R)^{\alpha-1}n(\R')^\beta+\beta n(\R')^\alpha 
n(\R)^{\beta-1} \}.
\label{aa3}
\end{equation}
Finally, we obtain
\begin{equation}
\frac{\delta^2 J}{\delta n(\R)\delta n(\R')}\arrowvert_{n=n_0}=2\alpha\beta n_0^{\alpha+\beta-2}w(\R-\R').
\label{aa4}
\end{equation}
Eq. (\ref{aa4}) combined with Eqs. (\ref{eq5}) and (\ref{eq6}) give Eq. (\ref{eq8}).

  \renewcommand{\theequation}{B-\arabic{equation}}
  \setcounter{equation}{0}  
  \section*{APPENDIX B}  

For a given $\Delta$, a series expansion of $F^{GAP}$ for $\eta\rightarrow 0$ gives:
\begin{equation}
F^{GAP}\longrightarrow \frac{3\Delta^2}{16\eta^2}+\frac{9}{5}+
\frac{3}{175}\frac{175\Delta^2-192}{\Delta^2}\eta^2+...,
\label{bb1}
\end{equation}
Thus, for any system with $\Delta >0$ we have that $F^{GAP}\propto \Delta^2 \eta^{-2}$, which is the 
most relevant physical result. We recall that for 
semiconductors and insulators, the density response function behaves as 
\cite{pick1970microscopic,huang2010nonlocal}
\begin{equation}
-\frac{1}{\chi^{Semic.}(k)}  \underset{k\rightarrow 0}{\longrightarrow}    \frac{b}{k^2},
\label{bb2}
\end{equation}
with $b\ge 0$ being material-dependent. Note that in the jellium-with-gap model, $b$ is a function of 
the band gap $E_g$.

On the other hand, if we first perform a series expansion for  $\Delta\rightarrow 0$, and then
a series expansion for  $\eta\rightarrow 0$ we obtain:
\begin{eqnarray}
&& F^{GAP}\longrightarrow \left[1+\frac{1}{3}\eta^2+\frac{8}{45}\eta^4+...\right]+
\Delta \left[\frac{\pi}{8}\frac{1}{\eta}+\frac{\pi}{12}\eta+...
\right]+...
\nonumber\\
&& F^{GAP}= F^{Lind}, \;\;\rm{when}\;\;\Delta=0\ .
\label{bb3}
\end{eqnarray}
such that at small band gaps, $F^{GAP}$ is close to the Lindhard function $F^{Lind}$
\begin{equation}
F^{GAP}\rightarrow F^{Lind}+\mathcal{O}(\Delta)+..., \;\rm{for}\;\Delta\rightarrow 0.
\label{bb4}
\end{equation}

In the limit of large wavevectors, i.e. for $\eta\rightarrow \infty$, we have
\begin{eqnarray}
&& F^{GAP}\rightarrow
3\eta^2-\frac{3}{5}+
 (-\frac{24}{175}+\frac{3}{16}\Delta^2)\frac{1}{\eta^2}+
\mathcal{O}(\frac{1}{\eta^4}) \,. 
\label{bb5}
\end{eqnarray}
Therefore, in this limit, $F^{GAP}$ always behaves as $F^{Lind}$ for $\Delta=0$.

Moreover, for any $\Delta$ and $\eta$, the following inequality holds (see Fig. 2 of Ref. 
\cite{PRBLGAP}).
\begin{equation}
F^{GAP}\ge F^{Lind}.
\label{bb6}
\end{equation}

  \renewcommand{\theequation}{C-\arabic{equation}}
  \setcounter{equation}{0}  
  \section*{APPENDIX C}  
Following Ref. \cite{wang1998orbital}, we can write Eq. (\ref{eq5}) in momentum space as 
\begin{eqnarray}
&& T_s[n]=\Omega \sum_\Q \tilde{t}_s^{\alpha,\beta}(\Q), \nonumber\\
&& \tilde{t}_s^{\alpha,\beta}(\Q)=\lambda 
\tilde{t}_{TF}(\Q)+\mu\tilde{t}_{W}(\Q)+\tilde{t}_X^{\alpha,\beta}(\Q), \nonumber\\
&&
\tilde{t}_{TF}(\Q)=\frac{3}{10}(3\pi^2)^{2/3}n_\Q^{5/6}n_{-\Q}^{5/6}, \nonumber\\
&&
\tilde{t}_{W}(\Q)=\frac{1}{2}n_\Q^{1/2}q^2n_{-\Q}^{1/2}
\label{cc1}
\end{eqnarray}
%
Let consider the partition (see also Ref. \cite{wang1998orbital})
\begin{equation}
\tilde{t}_X^{\alpha,\beta}(\Q)=-t_I(\Q)-t_{II}(\Q)-t_{III}(\Q),
\label{cc2}
\end{equation}
where 
\begin{eqnarray}
&& t_I(\Q)=\frac{1}{2\alpha\beta n_0^{\alpha+\beta-2}}n_\Q^\alpha \frac{1}{\chi^{GAP}}n_{-\Q}^\beta, 
\nonumber\\
&& t_{II}(\Q)=\lambda\frac{k_F^2}{6\alpha\beta n_0^{\alpha+\beta-1}}n_\Q^\alpha n_{-\Q}^\beta, \nonumber\\
&& t_{III}(\Q)=\mu \frac{1}{8\alpha\beta n_0^{\alpha+\beta-1}}n_\Q^\alpha q^2 n_{-\Q}^\beta.
\label{cc3}
\end{eqnarray}
Note that, for simplicity of notation, we use $n_\Q^\alpha G n_{-\Q}^\beta$ instead of the symmetric 
function $\frac{1}{2}\{ n_\Q^\alpha G n_{-\Q}^\beta +n_\Q^\beta G n_{-\Q}^\alpha \}$.

From Appendix B, we find 
\begin{eqnarray}
&& \lim_{\Q\rightarrow 0}\lim_{\Delta\rightarrow 
0}\frac{1}{\chi^{GAP}}=-\frac{1}{3n_0}(k_F^2+\frac{q^2}{12}), 
\nonumber\\
&& \lim_{\Q\rightarrow \infty}\frac{1}{\chi^{GAP}}=\frac{1}{n_0}(\frac{k_F^2}{5}-\frac{q^2}{4}),
\label{cc4}
\end{eqnarray}
%
then, substituting Eq. (\ref{cc4}) into Eq. (\ref{cc3}), we find after some algebra
\begin{eqnarray}
&&  \lim_{\Q\rightarrow 0}t_I(\Q)=-\frac{1}{\lambda}t_{II}(\Q)-\frac{1}{9\mu}t_{III}(\Q), \nonumber\\
&& \lim_{\Q\rightarrow 
0}\tilde{t}_s(\Q)=\lambda\tilde{t}_{TF}(\Q)+\mu\tilde{t}_{W}(\Q)+t_{II}(\Q)(\frac{1}{\lambda}-1)+ \nonumber\\
&& t_{III}(\Q)(\frac{1}{9\mu}-1), \nonumber\\
&& \lim_{\Q\rightarrow \infty}t_I(\Q)=\frac{3}{5\lambda}t_{II}(\Q)-\frac{1}{\mu}t_{III}(\Q), \nonumber\\
&& \lim_{\Q\rightarrow 
\infty}\tilde{t}_s(\Q)=\lambda\tilde{t}_{TF}(\Q)+\mu\tilde{t}_{W}(\Q)-t_{II}(\Q)(\frac{3}{5\lambda}+1)+ 
\nonumber\\
&& t_{III}(\Q)(\frac{1}{\mu}-1).
\label{cc5}
\end{eqnarray}
Performing the integrals, we find
\begin{eqnarray}
&& T_{III}=\Omega \sum_\Q \mu \frac{1}{8\alpha\beta n_0^{\alpha+\beta-1}}n_\Q^\alpha q^2 n_{-\Q}^\beta= 
\nonumber\\
&& \mu T_s^W +\mu (\alpha+\beta-1) \{ <\delta n|t_W>+ \frac{(\alpha+\beta-2)}{2}<\delta^2 n|t_W> 
\},\nonumber\\
\label{cc6}
\end{eqnarray}
and
\begin{eqnarray}
&& T_{II}=\lambda\frac{5}{9\alpha\beta}T_s^{TF}+\lambda\frac{5}{9\alpha\beta}(\alpha+\beta-\frac{5}{3})\times 
\nonumber\\
&& \{ <\delta n|t_{TF}>+\frac{1}{2}(\alpha+\beta-\frac{8}{3})<\delta^2 n|t_{TF}>   \}
\label{cc7}
\end{eqnarray}
Combining Eqs. (\ref{cc5})-(\ref{cc7}), we obtain Eqs. (\ref{eq9}) and (\ref{eq11}).

\twocolumngrid
\bibliography{kgap}
\bibliographystyle{apsrev4-1}

\end{document}